\documentclass[showpacs,twocolumn,prhrenumbers,amsmath,amssymb]{revtex4}
\usepackage{amsmath,amsfonts,latexsym,amssymb,graphicx,graphics,epsfig,subfigure,color,makeidx,xcolor,epstopdf,multirow}
\usepackage[colorlinks=true,citecolor=blue,linkcolor=red,anchorcolor=green,urlcolor=cyan]{hyperref}
\usepackage[title]{appendix}
\newcommand {\nn}{\nonumber}

\newcommand {\be}{\begin{equation}}
\newcommand {\ee}{\end{equation}}
\newcommand {\beq}{\begin{eqnarray}}
\newcommand {\eeq}{\end{eqnarray}}

\graphicspath{{plots/}}
\makeatletter

\begin{document}
\title{The characteristics of circular motion and quasiperiodic oscillations around accelerating black hole}

\author{Tao-Tao Sui$^{a}$\footnote{suitaotao@aust.edu.cn}}
\author{Xin-Yang Wang$^{a}$\footnote{wangxy@aust.edu.cn}}

\affiliation{$^{a}$Center for Fundamental Physics, School of Mechanics and Photoelectric Physics, Anhui University of Science and Technology, Huainan, Anhui 232001,  China\\
}

\begin{abstract}
{This study explores the motion of massive test particles and associated quasi-periodic oscillations (QPOs) around an accelerating black hole. The acceleration factor $A$ suppresses the radial effective potential $V_{\text{eff}}$, thereby lowering the energy $\mathcal{E}$ and angular momentum $\mathcal{L}$ required for stable circular orbits. Stability demands $\partial_r^2 V_{\text{eff}} \geq 0$, setting an upper bound $AM$ $\leq 0.0161$. As $A$ increases, the innermost stable circular orbit (ISCO) radius grows, while $\mathcal{L}{\text{ISCO}}$ and $\mathcal{E}{\text{ISCO}}$ decrease. Radiative efficiency $\epsilon$ rises with $A$, peaking at $6.9\%$. Fundamental frequencies show that $A$ accelerates the decay of the Keplerian $\Omega_{\phi}$ and vertical $\Omega_{\theta}$ frequencies, while suppressing the radial frequency E. The divergence between $\Omega_{\theta}$ and $\Omega_{\phi}$ increases with $A$, differing from spherical black hole behavior. Using the RP, ER3, ER4, and WD QPO models, the WD model predicts the highest frequencies. ER4’s resonant radius remains fixed across frequency ratios, unlike ER3. Although $A$ suppresses twin-peak QPO frequencies, it enhances the nodal precession frequency $\nu_{\text{nod}}$. Fitting observational data from GRO J1655-40 and XTE J1859+226 and applying the TOV limit, the ER4 model uniquely fits GRO J1655-40 with $(10^3A, M, r/M) \approx (4.31, 3.43 M_\odot, 8.08)$. For XTE J1859+226, three models yield $10^3A \approx 1.4$, excluding ER3, suggesting stronger acceleration in GRO J1655-40.}
\end{abstract}

\maketitle

\section{Introduction}\label{firstpart}
When a particle moves through a gravitational field, the field's characteristics are reflected in the particle's kinematic behavior. By analyzing the motion of test particles, a wide array of fascinating phenomena can be uncovered, including black hole shadows \cite{hennigar2018,EventHorizonTelescope:2019ths,Gao:2023mjb}, gravitational lensing \cite{Darwin59,Gyulchev:2008ff,Shaikh:2019jfr}, and the existence of stable circular orbits \cite{Zhang:2017nhl,Zhang:2021xhp,Yang:2021chw,Tan:2024hzw}. These features are common not only to black holes \cite{Virbhadra:2002ju,Joshi:2020tlq,Yang:2020jno}, but also to ultra-compact objects \cite{Cunha:2018acu,Shaikh:2019itn}.

The study of test particle geodesics around massive objects provides critical insights into the nature of these objects and the fundamental principles of gravity. In particular, the unique features of black holes become especially evident in the strong-field regime, where their properties infulence the radiation emitted by infalling or orbiting matter. Numerous studies have explored the dynamics of particles around rotating black holes \cite{Wilkins:1972rs,Glampedakis:2002ya,Fujita:2009bp,Pugliese:2013zma}, as well as the behavior of circular motion on the equatorial plane of distorted, spherically symmetric black holes \cite{Shoom:2015slu,Faraji:2020skq}.

The motion of test particles is central to astrophysical investigations of black holes. Although radiation cannot escape from within the event horizon, emissions from the surrounding region, particularly from accretion disks—offer vital information about the near-horizon geometry \cite{Bardeen:1972fi}. Accretion disks emit soft X-ray continuum radiation, the frequency of this radiation can be used to estimate their inner edge of the disk. This inner radius typically corresponds to the innermost stable circular orbit (ISCO), which encodes essential information about the central compact object. Consequently, the dynamics of circular orbits govern the behavior of accretion disks and reflect fundamental properties of the black hole spacetime.
Moreover, accretion disks are the primary sources of electromagnetic radiation in black hole systems \cite{Narayan:1993bd,Tanaka:1995en}, making them natural laboratories for probing matter and gravity under extreme conditions \cite{Kato:2022cur,Kadowaki:2018dtb,Andreoni:2022afu,Kosec:2023qva}.

As early as the 1970s, Syunyaev and collaborators \cite{Syunyaev1972} proposed probing strong gravitational fields by analyzing rapid variability in of X-ray flux emitted from the innermost regions of accretion disks. This idea gained renewed significance following the discovery of quasi-periodic oscillations (QPOs) in the X-ray emissions from accreting compact objects. These astrophysical signals, with frequencies as high as 450 Hz, are consistent with those expected from bound orbits near the innermost stable circular orbit (ISCO) {\cite{Lewinbook, Motta:2016vwf,Rezzolla:2003zx,Montero:2011nq}}. Observational evidence further suggest the presence of QPOs in the continuous X-ray emissions of (micro)quasars—binary systems composed of a black hole or neutron star and a companion star.

QPOs span a wide frequency range, from a few millihertz to approximately 0.5 kHz, and are typically classified into high-frequency (HF, $0.1\text{–}1$ kHz) and low-frequency (LF, $<0.1$ kHz) categories \cite{Stella:1997tc, Stella:1999sj}. HF QPOs are often observed as twin peaks, representing upper and lower frequency components with characteristic frequency ratios often close to $3:2$ in black hole microquasars \cite{Kluzniak:2001ar,Abramowicz:2002xc,Abramowicz:2004rr,Aliev:1980hz}. These oscillations serve as powerful probes of strong-field gravity and offer insights into the structure of inner accretion disks, as well as the mass, radius, and spin of compact objects including white dwarfs, neutron stars, and black holes. As a result, a range of theoretical models has been developed to explain the origin of HF QPOs, which in turn impose constraints on the physical parameters of compact sources \cite{Stuchlik:2013esa,Maselli:2014fca,Jusufi:2020odz,Ghasemi-Nodehi:2020oiz,Chen:2021jgj,Allahyari:2021bsq,Deligianni:2021ecz,Deligianni:2021hwt,Jiang:2021ajk,Banerjee:2022chn,Liu:2023vfh,Riaz:2023yde,Rayimbaev:2023bjs,Abdulkhamidov:2024lvp,Jumaniyozov:2024eah,Guo:2025zca,Zhang:2025acq}. Future high-precision observations with instruments such as Insight-HXMT (Hard X-ray Modulation Telescope) \cite{Lu:2019rru} and the upcoming Einstein Probe \cite{yuan2018einstein} are expected to place stringent constraints on the physical parameters of these compact sources. {Besides, Ref. \cite{Shahzadi:2023act} maps a set of astrophysically motivated deviations from the classical Kerr black hole spacetime and identifies those that best fit the HF QPOs data.}

In parallel with observational advances, novel mechanisms for black hole formation have been explored. Notably, black holes may be produced in pairs in the presence of cosmic strings \cite{hr,eardley1995,ashoorioon2014}, in de Sitter backgrounds \cite{mellor1989,mann1995,dias2004:2}, or under the influence of external magnetic fields \cite{garfinkle1991,hawking1995,dowker1994}. Alternatively, primordial black holes formed in the early universe may become trapped within cosmic string networks \cite{Vilenkin:2018zol}. In these scenarios, the tension of the cosmic string induces an acceleration of the black hole. Such configurations are described by a particular class of solutions to Einstein’s field equations known as the C-metric \cite{Weyl:1917rtf,Robinson:1962zz,Griffiths:2006tk}, which represents black holes undergoing constant acceleration due to a conical deficit along their polar axis \cite{kinnersley1970}. Notably, even small accelerations may have non-negligible effects on galactic structure formation \cite{vilenkin2018,gussmann2021,Morris_2017}. 

The presence of acceleration leads to distinct observational signatures. For instance, the time delays associated with gravitational lensing by accelerating black holes differ significantly from those of non-accelerating ones \cite{Ashoorioon:2022zgu,JahaniPoshteh:2022yei}. Additionally, the optimal viewing angle for observing shadows of accelerating black holes deviates from the equatorial plane ($\theta = \pi/2$) \cite{Grenzebach_2015,Zhang:2020xub,EslamPanah:2024dfq,Sui:2023rfh}. Motivated by these considerations, this work investigates the epicyclic motion of massive test particles around an accelerating Schwarzschild black hole and analyzes how the acceleration parameter influences their kinematic properties. We further examine the fundamental frequencies of these particles' orbits, which are directly linked to QPO phenomena. Specifically, we explore the impact the influence of the acceleration factor on four prominent models of twin-peak HF QPOs and fit the resulting theoretical predictions to the observed X-ray QPO frequencies of GRO J1655-40 and XTE J1859+226. {This analysis can serve as supplementary content to the section in \cite{Shahzadi:2023act} that discusses the discrepancies in the characteristics of QPOs between accelerating Kerr black hole and classical Kerr black hole, thereby enhancing our understanding of their astrophysical manifestations.}

The paper is organized as follows: In Section \ref{secondpart}, we will analyze the circular motion of test particle around accelerating black holes.  Section \ref{thirdpart} focuses on the calculation of the fundamental frequencies of test particle orbits, including the Keplerian fre- quency and the radial and vertical epicyclic frequencies relative to the circular trajectory.  
In Section \ref{fourthpart}, we consider the characteristic frequencies predicted by various twin-peak QPO models and constrain the corresponding parameters using observational data from the two microquasars. Finally, in Section \ref{conclusion}, we summarize our key findings and present concluding remarks.

\section{Circular motion of test particle around accelerating black holes}\label{secondpart}
In this section, we will precisely review the dynamics of test particles around an accelerating Schwarzschild black holes which can be described by C-metric. The metric can be written in the following form \cite{griffiths2006new}
\begin{eqnarray}
ds^2=\frac{1}{\Omega^2}\Big[-\Delta_rdt^2+\frac{dr^2}{\Delta_{r}}+\frac{r^2d\theta^2}{P_\theta}+P_\theta r^2\sin^2\theta d\phi^2\Big],\label{eqn:metric:griffiths}
\end{eqnarray}
with 
\begin{eqnarray}
&&\Omega=1+Ar\cos(\theta),~P_\theta=1+2AM\cos(\theta), \nn\\ 
&&\Delta_r=(1-A^2r^2)\left(1-2M/r\right),
\label{eqn:tran:griffiths}
\end{eqnarray} 
where $M$ and $A$ denote the mass parameter and acceleration factor of black hole, respectively. From the conformal factor $\Omega$, we see that the boundary is located at $r_{\text{bd}}=1/|A\cos(\theta)|$. To ensure the black hole is contained within the spacetime (i.e., inside the conformal boundary), the event horizon at $r_{\text{eh}}=2M$ must satisfy $r_{\text{eh}}\gg1/A $, which implies the condition $AM \ll 0.5$.

\subsection{Equation of motion for test particles}
For a test particle with unit mass, the equation of motion can be derived by the Lagrangian density as 
\begin{equation}
\mathcal{L}_{Ld}=\frac{1}{2}g_{\mu\nu}\dot{x}^\mu\dot{x}^\nu, 
\end{equation}
with the definition $\dot{x}^\mu=dx^\mu/\lambda$, where $\dot{x}^\mu$ is the four-velocity of the test particles and the parameter $\lambda$ is the affine parameter. For this stationary black hole, there are two Killing vectors $\partial_t$ and $\partial_\phi$ which can result the conserved total energy $\mathcal{E}$ and and z-component of the angular momentum $\mathcal{L}$ as
\begin{equation}
\mathcal{E}=-g_{tt}\dot{t},~~\mathcal{L}=g_{\phi\phi}\dot{\phi}.
\end{equation}
Using the normalization condition for the four-velocity, $g_{\mu\nu}\dot{x}^\mu\dot{x}^\nu=-1$, we can write the motion equation as
\begin{eqnarray}
&&g_{rr}\dot{r}^2+g_{\theta\theta}\dot{\theta}^2=-1+\frac{\mathcal{E}^2}{g_{tt}}-\frac{\mathcal{L}^2}{g_{\phi\phi}}=V_{\text{all}},\nn\\
&&V_{\text{all}}=-1+\frac{\mathcal{E}^2\Omega^2}{\Delta_r}-\frac{\mathcal{L}^2\Omega^2}{P_\theta r^2\sin(\theta)^2}.\label{eomall}
\end{eqnarray}
The full equations of motion in each coordinate direction are 
\begin{eqnarray}
&&\dot{t}=\frac{\mathcal{E}\Omega^2}{\Delta_r},~~~~~~~~\dot{\phi}=\frac{\mathcal{L}\Omega^2}{P_\theta r^2 \sin(\theta)^2},\\ 
&&\frac{\dot{r}^2}{\Omega^2}=\mathcal{E}^2\Omega^2-\Delta_r\Big(1+\frac{\mathcal{K}}{r^2}\Big),\\
&&\frac{\dot{\theta}^2}{\Omega^2}=\frac{P_\theta}{r^4}\Big(\mathcal{K}-\frac{\mathcal{L}^2\Omega^2}{P_\theta \sin(\theta)^2} \Big)=V_{\theta},\label{eomtheta}
\end{eqnarray} 
where $\mathcal{K}$ is the Carter constant. 

\begin{figure}[htbp!]
\centering 
\includegraphics[width=0.4\textwidth]{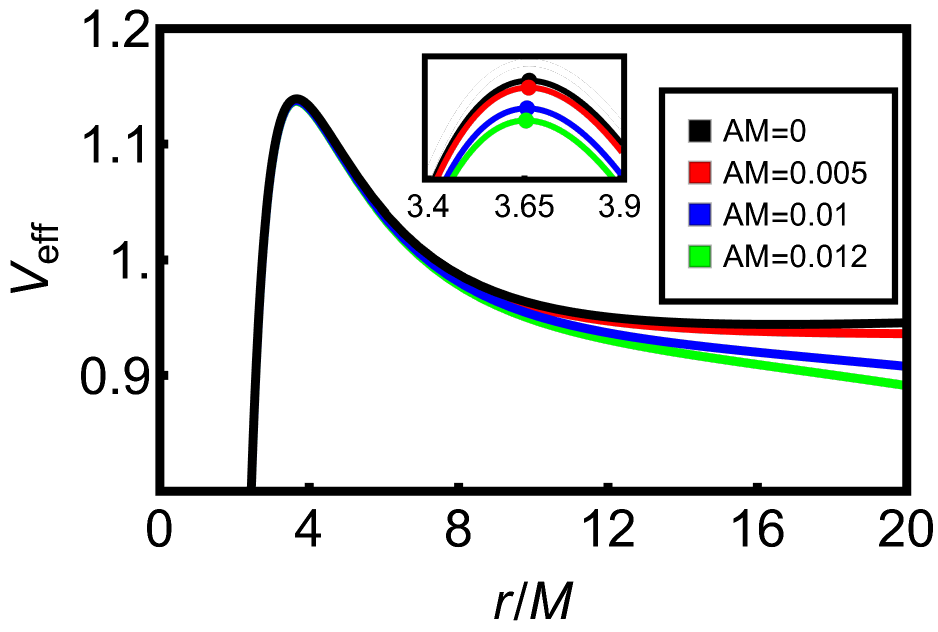}
\caption{Effective potentials $V_{\text{eff}}$ for the radial motion of test particles with angular momentum $\mathcal{L}=4.5M$, with different values of acceleration parameter $AM$.}\label{effpotential}
\end{figure}

In the remainder of this work, we restrict our analysis to equatorial motion, i.e., $\theta=\pi/2$ and $\dot{\theta}=0$, which implies $\mathcal{K}=\mathcal{L}^2$. Under this condition, the radial equation of motion simplifies to:
\begin{eqnarray} \label{eomr}
&&\dot{r}^2=\mathcal{E}^2-l\Delta_r\Big(1+\frac{\mathcal{L}^2}{r^2}\Big)=\mathcal{E}^2-V_{\text{eff}},\\
&&V_{\text{eff}}=\Delta_r\Big(1+\frac{\mathcal{L}^2}{r^2}\Big),\nn
\end{eqnarray} 
Here, $V_{\text{eff}}$ acts as the effective potential for radial motion. Figure~\ref{effpotential} illustrates how the acceleration factor $A$ influences the effective potential. From As shown, increasing the value of $A$ lowers the overall height of the effective potential. Additionally, the peak of $V_{\text{eff}}$ shifts to smaller radial values as $A$ increases. This behavior indicates that the acceleration parameter significantly affects other kinematic properties of particle motion, which we will explore in the following sections.

\subsection{Circular orbits}
For the circular orbits of the massive test particles around the accelerating Schwarzschild black hole, the equation of motion in the radial direction should satisfy the below conditions as 
\begin{equation}
\dot{r}=0,~~\ddot{r}=0 ~\Longrightarrow~ V_{\text{eff}}=\mathcal{E}^2,~~\partial_{r}V_{\text{eff}}=0. \label{circular condition}
\end{equation}
By solving the above two condition equations, we can obtain the explicit relations between the angular momentum $\mathcal{L}$ (the total energy $\mathcal{E}$) and radius $r$ of the  circular orbits, 
\begin{eqnarray}
&&\mathcal{L}=\sqrt{\frac{r^2\big(A^2 M r^2-A^2 r^3+M\big)}{r+M \big(A^2 r^2-3\big)}},\\
&&\mathcal{E}=\sqrt{\frac{\big(A^2 r^2-1\big)^2(r-2 M)^2}{r+M \big(A^2 r^2-3\big)}}.
 \label{circularel}
\end{eqnarray}

The radial dependence of total energy $\mathcal{E}$ and angular momentum $\mathcal{L}$ for the test particles on circular orbits around the accelerating black hole is shown in Fig. \ref{circularel}. As shown in Fig. \ref{circularel}, both $\mathcal{E}$ and $\mathcal{L}$ decrease with increasing acceleration factor $A$. Furthermore, the presence of acceleration lowers the minimum values of both energy and angular momentum, while shifting the corresponding orbital radius outward. This critical radius can be interpreted as the radius of the innermost stable circular orbit (ISCO). Consequently, the ISCO radius increases as $A$ becomes larger.

\begin{figure}[htp!]
\includegraphics[width=0.4\textwidth]{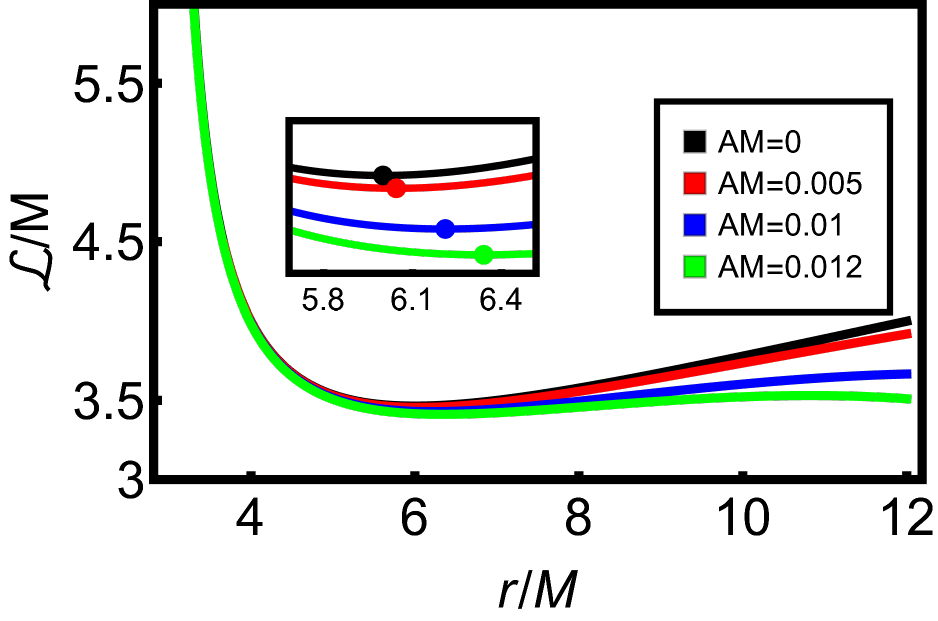}
\includegraphics[width=0.4\textwidth]{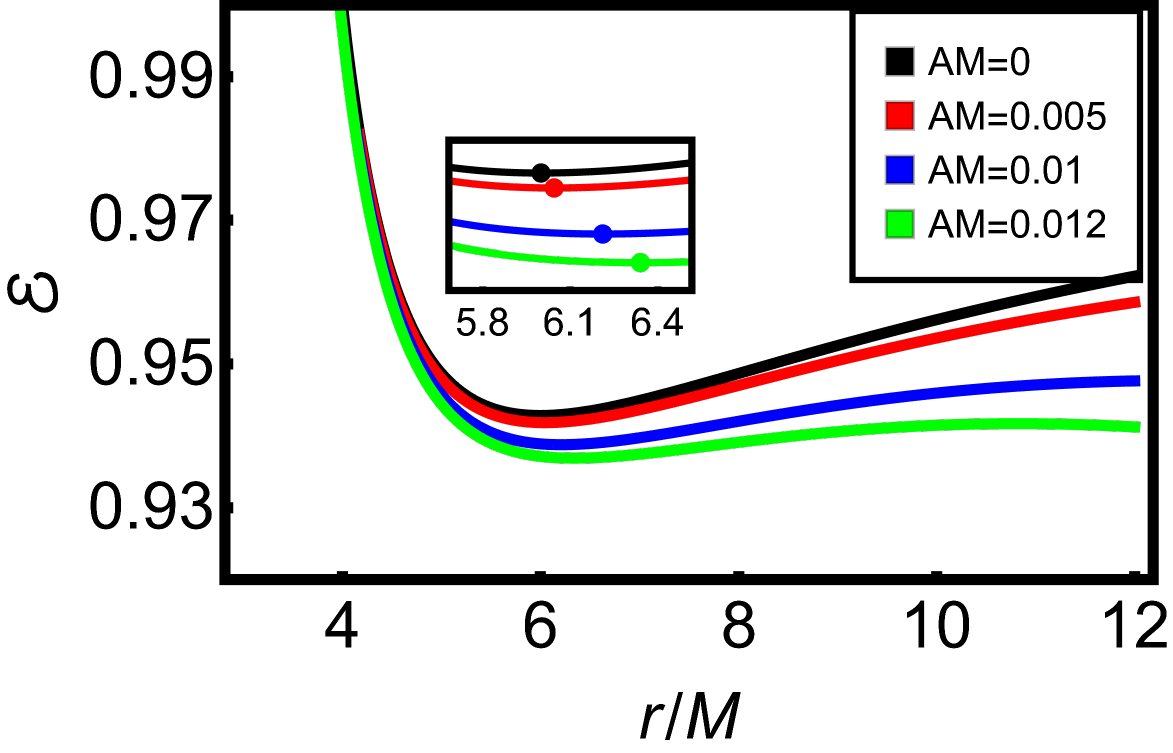}
\caption{The influence of the acceleration factor $A$ on the angular momentum $\mathcal{L}$ and specific energy $\mathcal{E}$ of test particles in circular orbits around accelerating black hole.}
\label{circularel}
\end{figure}

\subsection{Innermost stable circular orbits}
In general, not all the circular orbits are stable. To identify the stable circular orbits, the second derivative of the effective potential must satisfy the condition: $\partial_r^2 V_{\text{eff}}\geq0$, or equivalently,
\begin{equation}
2r \Delta_r\partial^2_r\Delta_r-4 r (\partial_r\Delta_r)^2+6\Delta_r\partial_r\Delta_r\geq0.\label{iscocondition}
\end{equation}
This requirement places a constraint on the acceleration factor, which results in an upper bound of  $AM\leq0.0161$. 

\begin{figure}[htbp!]
{\includegraphics[width=0.4\textwidth]{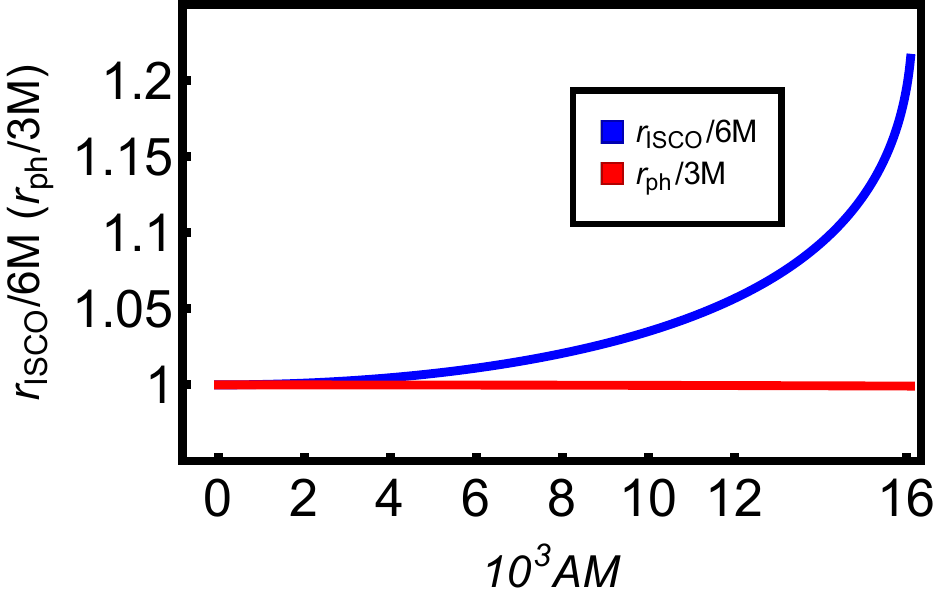}}
\caption{Ratio of the ISCO and photon sphere radii of an accelerating black hole to those of a Schwarzschild black hole, with different values of the acceleration parameter $AM$.}\label{iscor}
\end{figure}

\begin{figure}[htp!]
\includegraphics[width=0.4\textwidth]{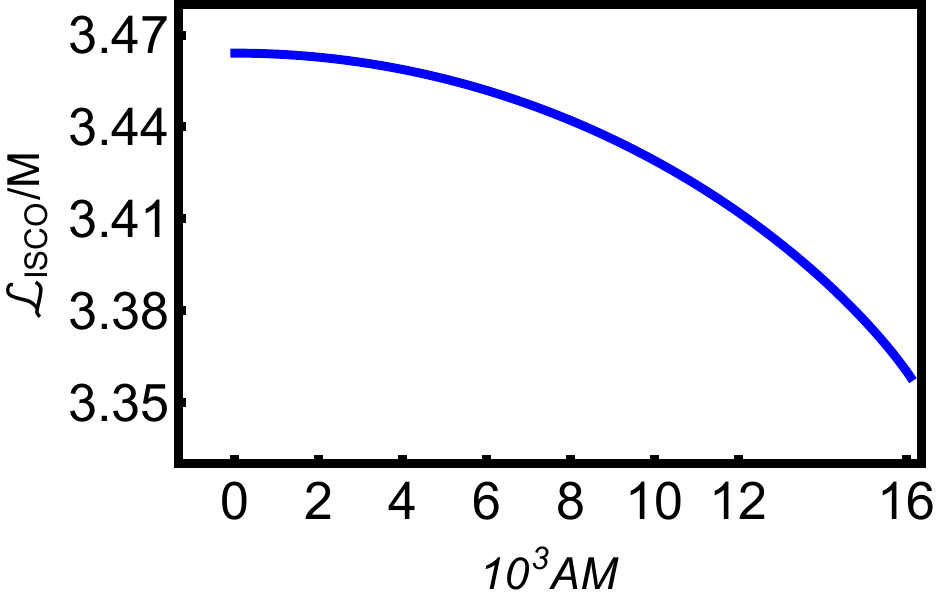}
\includegraphics[width=0.4\textwidth]{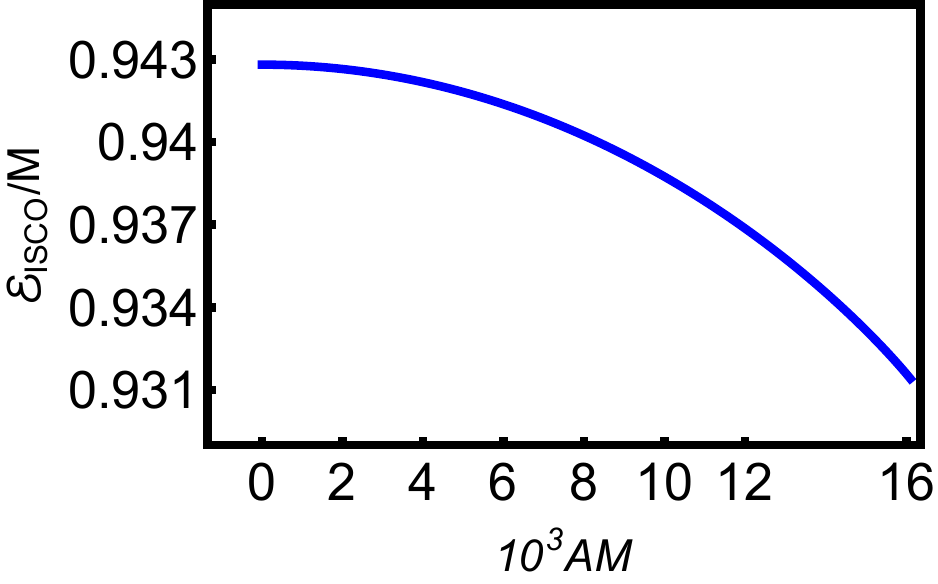}
\caption{The variation of the angular momentum  $\mathcal{L}_{ISCO}$ and total energy $\mathcal{E}_{ISCO}$ at the ISCO with respect to the acceleration parameter $AM$.}
\label{iscole}
\end{figure}

By solving the equation $\partial^2_rV_{\text{eff}}=0$, we determine the location of the ISCO as a function of the acceleration factor $A$. Figure \ref{iscor} illustrates how the ISCO radius and the photon sphere radius evolve with increasing $A$, compared to the Schwarzschild black hole case. As shown, the ISCO radius increases monotonically with $A$, and the rate of increase also grows with larger values of $A$. In contrast, the acceleration parameter has a negligible effect on the photon sphere radius $r_{\text{ph}}$. Besides, Fig. \ref{iscole} shows that both the angular momentum $\mathcal{L}_{ISCO}$ and the total energy $\mathcal{E}_{ISCO}$ associated with the ISCO decrease as $A$ increases, a trend consistent with that observed in Fig. \ref{circularel}.  

\begin{figure}[htbp!]
{\includegraphics[width=0.4\textwidth]{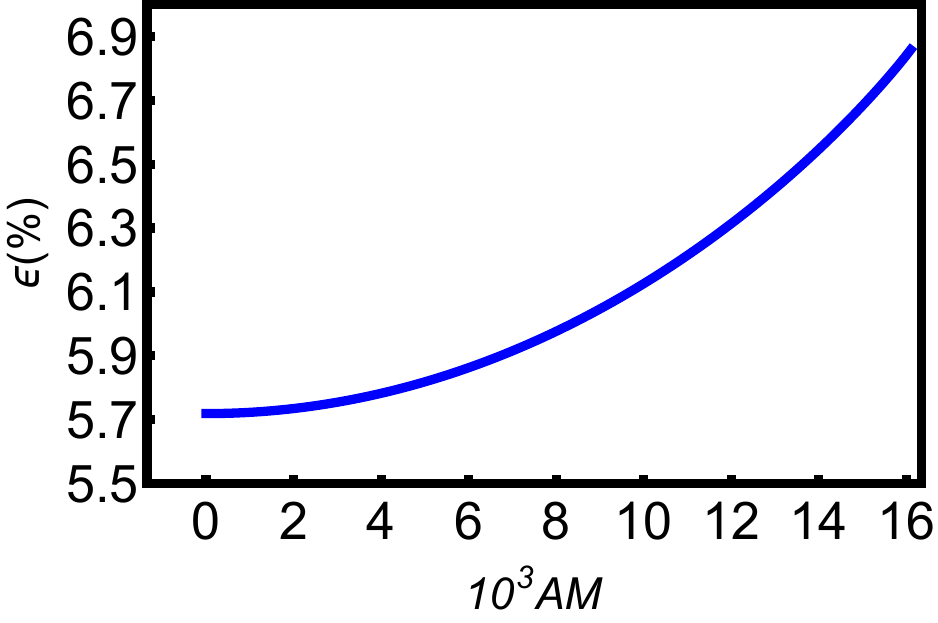}}
\caption{Radiative efficiency $\epsilon $ as a function of the acceleration parameter $\text{AM}$.}\label{iscoeff} 
\end{figure}

According to the thin accretion disk model, a test particle falling from infinity into a black hole radiates energy as it spirals inward \cite{Collodel:2021gxu,Wu:2024sng,Kurmanov:2024hpn,Liu:2024brf}. For a unit-mass test particle, the maximum radiative efficiency $\epsilon$ is defined as the difference between the rest energy and the orbital energy at  ISCO. It is given by:
\begin{equation}
\epsilon=1-\mathcal{E}_{ISCO}.
\end{equation}
Figure. \ref{iscoeff} shows how the radiative efficiency $\epsilon$ varies with the acceleration factor $A$ for an accelerating black hole. As depicted, the radiative efficiency $\epsilon$ increases monotonically with the acceleration factor $A$, reaching a maximum value of approximately $6.9\%$. Moreover, the rate of growth in radiative efficiency $\epsilon$ also becomes steeper as factor $A$ grows. This indicates that the acceleration factor $A$ not only enhances the motion of the black hole but also significantly boosts the energy released from infalling test particles during accretion.

\section{Fundamental frequencies}\label{thirdpart}
In this section, we investigate the fundamental frequencies associated with test particles orbiting an accelerating black hole. In particular, we focus on  the influence of the acceleration factor $A$ on the Keplerian orbital frequency, the radial and vertical epicyclic frequencies, which characterize oscillations around stable circular orbits.

\subsection{Keplerian frequency}
\begin{figure}[htbp!]
{\includegraphics[width=0.4\textwidth]{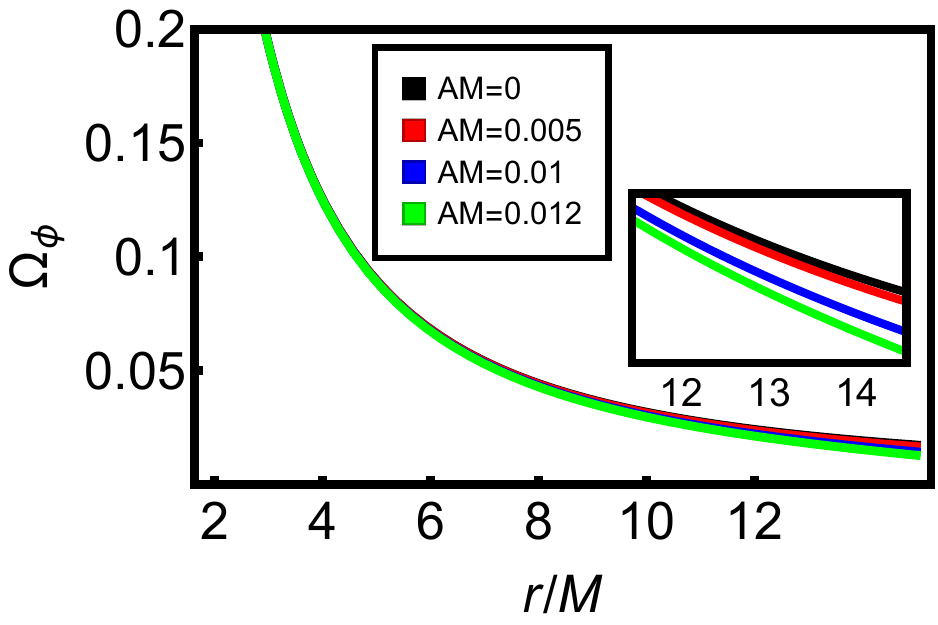}}
\caption{Keplerian frequency $\Omega_\phi$ of test particles orbiting an accelerating black hole as a function of the radial coordinate, shown for different values of the acceleration parameter $\text{AM}$.}\label{Kfrequency}
\end{figure}

For an observer at infinity, the angular velocity of test particles orbiting a black hole corresponds to the orbital (Keplerian) frequency $\Omega_\phi$, which is defined as $\Omega_\phi=\dot{\phi}/\dot{t}$. In the equatorial plane, $\Omega_\phi$ can be expressed with
\begin{equation}
\Omega_\phi^2=\partial_r\Delta_r/2r=A^2\big(\frac{M}{r}-1\big)+\frac{M}{r^3},
\end{equation}
The radial dependence of the Keplerian frequency $\Omega_\phi$ for test particles for different values of the acceleration factor $A$ is shown in Fig. \ref{Kfrequency}. As illustrated, the presence of the acceleration factor $A$ enhances the decay rate of Keplerian frequency $\Omega_\phi$ with increasing radial coordinate $r/M$, indicating that acceleration suppresses the orbital frequency more rapidly at larger radii.

{Additionally, it is known that a nonmonotonic behavior of the angular velocity $\dot{\phi}/\dot{t}$ within the regime of stable circular orbits can give rise to the Aschenbach effect \cite{Aschenbach:2004kj}, which was firstly discovered and discussed for the case of near-extreme Kerr BH \cite{Stuchlik:2004wk} and spherically symmetric BH \cite{Wei:2023fkn}}. However, no such nonmonotonic behavior is observed in Fig. \ref{Kfrequency}. This implies that the Aschenbach effect does not occur for equatorial circular orbits around accelerating Schwarzschild black hole.

\subsection{Harmonic oscillations}
The radial and vertical fundamental frequencies of test particles on circular orbits in the equatorial plane can be derived by introducing small perturbations in the radial and vertical directions, namely, $r\rightarrow r_0+\delta r$ and $\theta\rightarrow \pi/2+\delta \theta$, respectively. Under the conditions for circular motion, $V_{\text{all}}(r_0,\frac{\pi}{2})=0$ and $\partial_{r(\theta)}V_{\text{all}}=0$, the motion in the perturbed directions can be approximated as harmonic oscillations. Therefore, the radial and vertical frequencies, as observed by a distant observer, are governed by the harmonic oscillator equations:
\begin{eqnarray}
\frac{d^2\delta \theta}{dt^2}+\Omega_\theta^2\delta \theta=0,~~\frac{d^2\delta r}{dt^2}+\Omega_r^2\delta r=0,\label{harmoniceq}
\end{eqnarray}
with 
\begin{eqnarray}
&&\Omega_\theta^2=-\frac{1}{2g_{\phi\phi}\dot{t}^2}\partial^2_{\theta}V_{\text{all}}|_{\theta=\frac{\pi}{2}},\label{hoscillatot}\\
&&\Omega_r^2=-\frac{1}{2g_{rr}\dot{t}^2}\partial^2_{r}V_{\text{all}}|_{\theta=\frac{\pi}{2}},\label{hoscillator}
\end{eqnarray}
the frequencies of the vertical and radial oscillations, respectively. After straightforward algebraic calculations, we can easily get the expressions for the vertical and radial  frequencies can be expressed as, 
\begin{eqnarray}
&&\Omega_\theta^2=\frac{\left(A^2 (r-2 M)^2+1\right)\partial_r \Delta _r}{2 r}-A^2 \Delta _r,\label{oscillatot}\\
&&\Omega_r^2=\frac{1}{2} \Delta _r\big(\frac{3\partial_r\Delta_r}{r}+\partial^2_r\Delta_r\big)-(\partial_r \Delta _r)^2.\label{oscillator}
\end{eqnarray}

\begin{figure}[htbp!]
\subfigure[$\Omega_\theta$]{\label{vertical oscillation}
\includegraphics[width=0.43\textwidth]{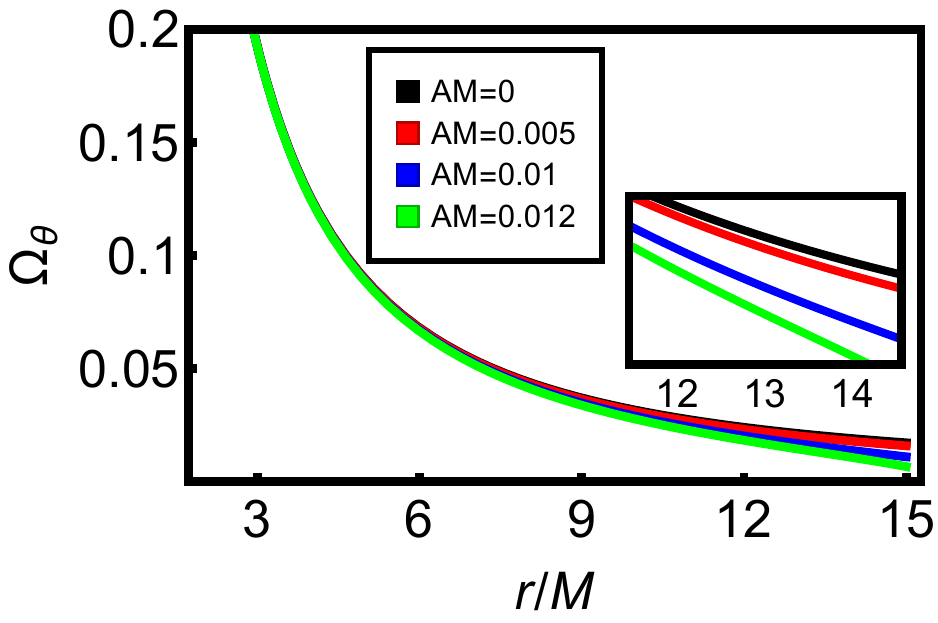}}
\subfigure[$10^3(\Omega_\phi-\Omega_\theta)$]{\label{deltaKV}
\includegraphics[width=0.4\textwidth]{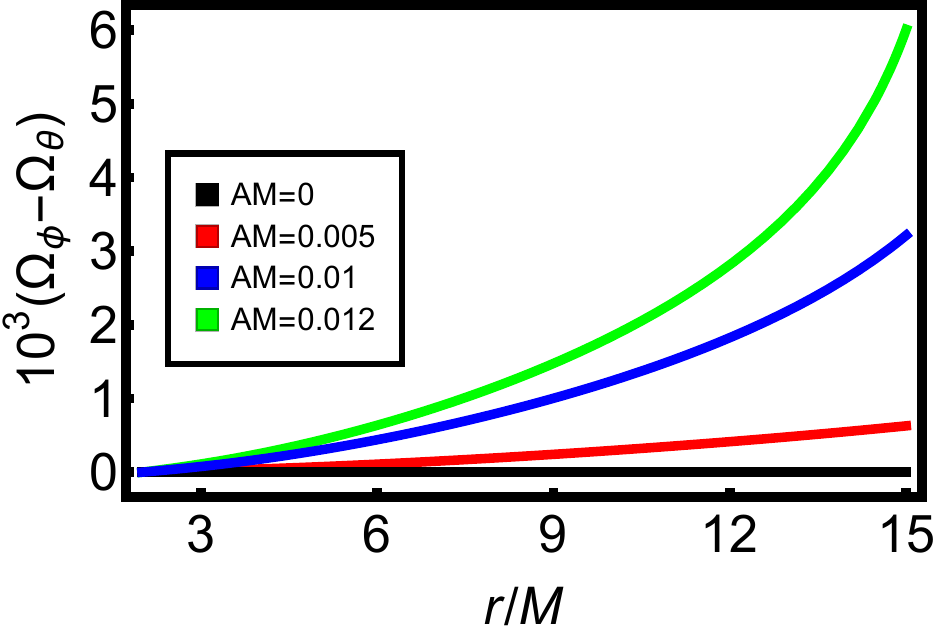}}
\caption{Vertical frequency $\Omega_\theta$ of massive test particles orbiting an accelerating black hole, and the difference between the vertical frequency $\Omega_\theta$ and the Keplerian frequency $\Omega_\phi$, i.e., $10^3(\Omega_\phi-\Omega_\theta)$, shown for various values of the acceleration parameter $AM$.}\label{vkv oscillation}
\end{figure}

Figure \ref{vertical oscillation} shows the radial dependence of the vertical frequency of the test particle orbiting an accelerating black hole. This frequency exhibits behavior qualitatively similar to that of Keplerian frequency $\Omega_\phi$. Besides, a comparison of the analytical expressions for $\Omega_{\phi}$ and $\Omega_{\theta}$ reveals that their difference increases with the acceleration factor $A$, as demonstrated in Fig. \ref{deltaKV}. {Furthermore, it is also imperative to note that for the charged particles, the disparity between the $\Omega_{\phi}$ and $\Omega_{\theta}$ cannot be overlooked for the cases of magnetized black hole (surrounded by magnetic field and observed commonly), and thoroughly discussed in Ref. \cite{Stuchlik:2020rls}.}

\begin{figure}[htbp!]
{\includegraphics[width=0.4\textwidth]{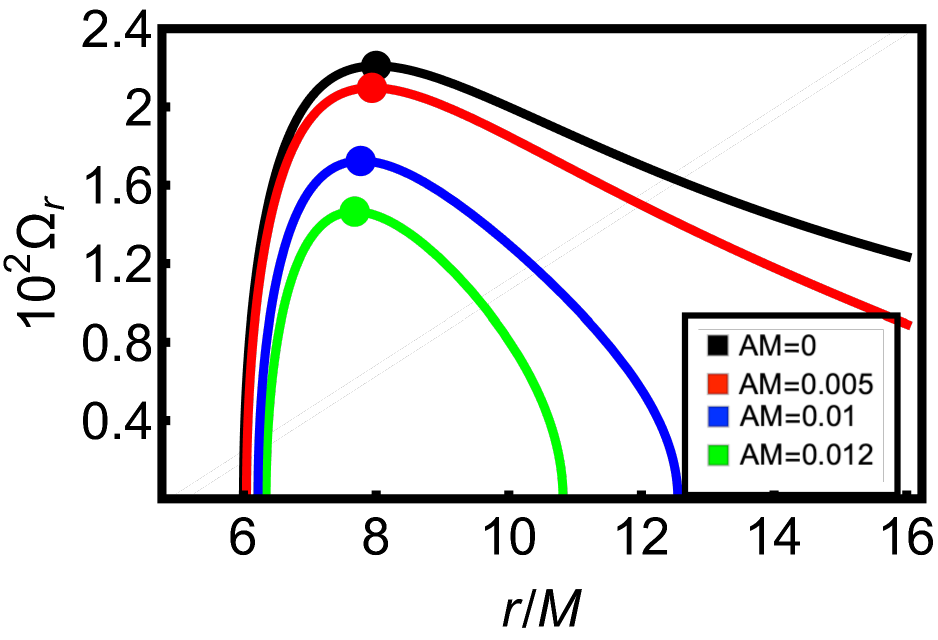}}
\caption{Radial dependence of the radial oscillation frequency $\Omega_r$ of test particles for different values of the acceleration parameter $AM$.}\label{radia oscillation}
\end{figure}

By combining the stability condition for circular orbits,  $\partial_r^2V_{\text{eff}}\geq0$, with the expression for the radial epicyclic frequency $\Omega_r^2$ (See Eq. \eqref{hoscillator}), it follows that radial harmonic oscillations are confined to the region of stable circular orbits—bounded by the innermost and outermost stable circular orbit radii. Figure \ref{radia oscillation} shows the influence of the acceleration factor $A$ on the radial frequency $\Omega_r^2$ of the test particles around accelerating black hole. From Fig. \ref{radia oscillation}, one can easily see that the maximum value of the radial frequency $\Omega_r|_{\text{max}}$ decreases with the acceleration factor $A$, as well as the distance where the radial frequency takes its maximum decreases with $A$. In addition, a larger acceleration factor $A$ leads to a narrower range of stable circular orbits, as clearly illustrated in Fig \ref{radia oscillation}.

\section{Quasi-periodic oscillation around accelerating black hole}\label{fourthpart}

In this section, we analyze the possible frequency predictions of various twin-peak QPO models in the context of an accelerating black hole. In most cases, these models highlight the essential role of the gravitational field of the central black hole, with the observed QPO frequencies closely tied to the fundamental frequencies of harmonic oscillations along geodesic orbits. For static observers at infinity, the fundamental frequencies in physical units [Hz] can be expressed as:
\begin{eqnarray}
\nu_{i}=\frac{1}{2\pi}\frac{c^3}{G M}\Omega_{i} [\text{Hz}],~~(i=r,\phi,\theta).
\end{eqnarray}  
We now examine the influence of the acceleration parameter $A$ on twin-peak QPOs around a central black hole, considering various QPO models and different values of the black hole’s physical parameters. In particular, we focus on the following models:

(a) The Relativistic Precession (RP) model, which is originally proposed to explain twin-peak QPOs in both black holes and neutron stars \cite{Stella:1999sj}, attributes the observed frequencies to relativistic precession effects. According to the RP model, the upper and lower QPO frequencies are identified as $\nu_U =\nu_{\phi}$ and $\nu_L=\nu_\phi-\nu_r$, respectively.

(b) The Epicyclic Resonance (ER) model, explores resonances arising from axisymmetric oscillation modes of thick accretion disk around {black holes \cite{Stuchlk:2016,Abramowicz:2001bi}}, which are associated with the circular geodesics of test particles. Here, we just consider two different kinds of ER mode, namely ER3 and ER4 mode. According to the ER model, the upper and lower frequencies can be defined with $\nu_U =\nu_{\theta}+\nu_{r}$, $\nu_L=\nu_\theta$ for ER3 mode and $\nu_U =\nu_{\theta}+\nu_{r}$, $\nu_L=\nu_\theta-\nu_r$ for ER4 mode, respectively.

(c) The Warped Disc (WD) model, built on the assumption of non-axisymmetric oscillatory modes within warped thin accretion disk surrounding black holes \cite{Kato:2004vs,Kato:2008sr}. In the WD model, the QPO frequencies of upper and lower are defined as $\nu_U =2\nu_{\phi}-\nu_{r}$, $\nu_L=2(\nu_\phi-\nu_r)$.

{Furthermore, Cadez et al. also introduced the Tidal Disruption (TD) model, which interprets QPOs as manifestations of the tidal disruption of large accreting inhomogeneities  \cite{Cadez:2008iv,Kostic:2009hp}, and more detailed exposition of QPOs models can be found in Refs. \cite{Kotrlova:2014ana,Kolos:2020ykz}. }

\begin{figure*}[!htb]
\begin{center}
\subfigure[RP model]{\label{RP model}
\includegraphics[width=0.4\textwidth]{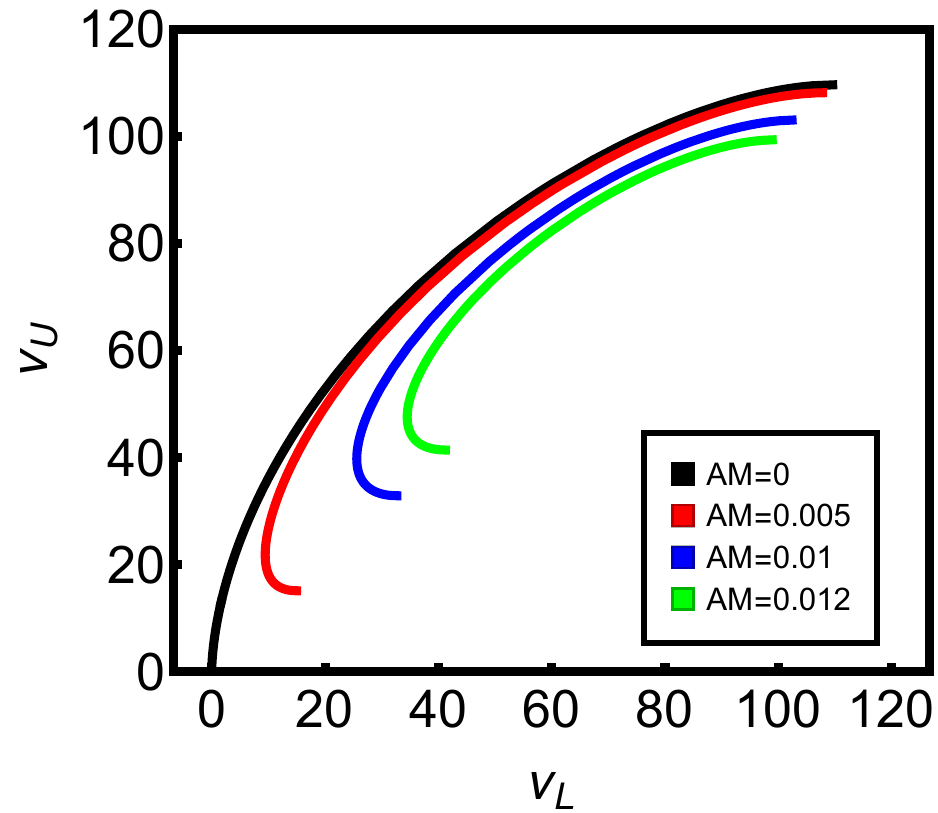}}
\subfigure[ER3 model]{\label{ER3 model}
\includegraphics[width=0.4\textwidth]{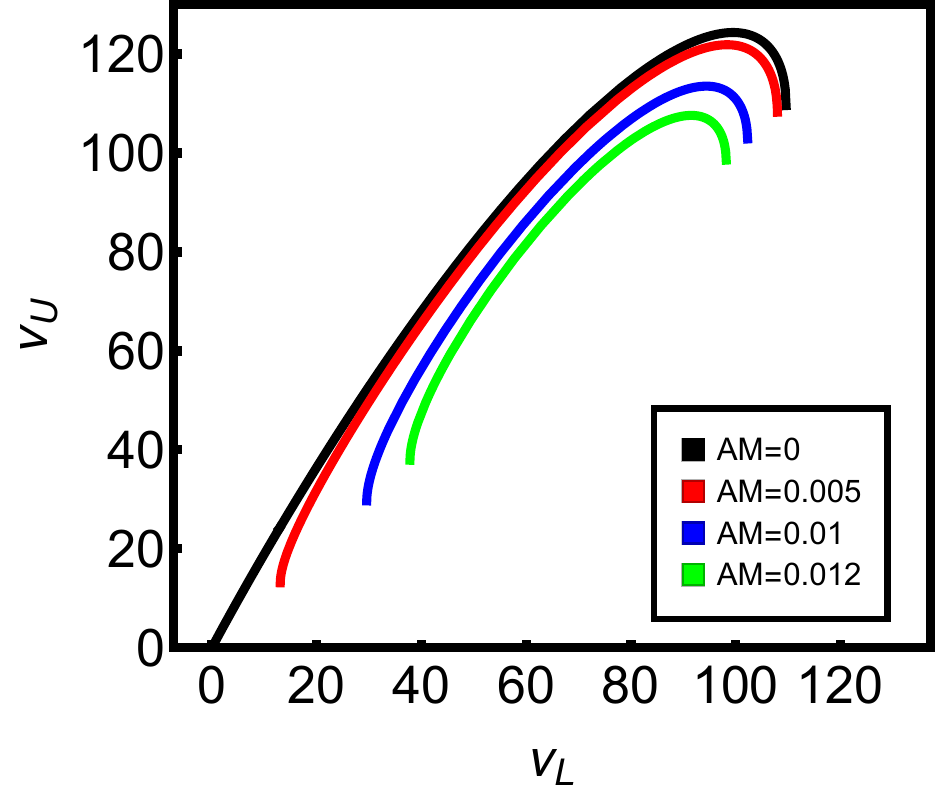}}
\subfigure[ER4 model]{\label{ER4 model}
\includegraphics[width=0.4\textwidth]{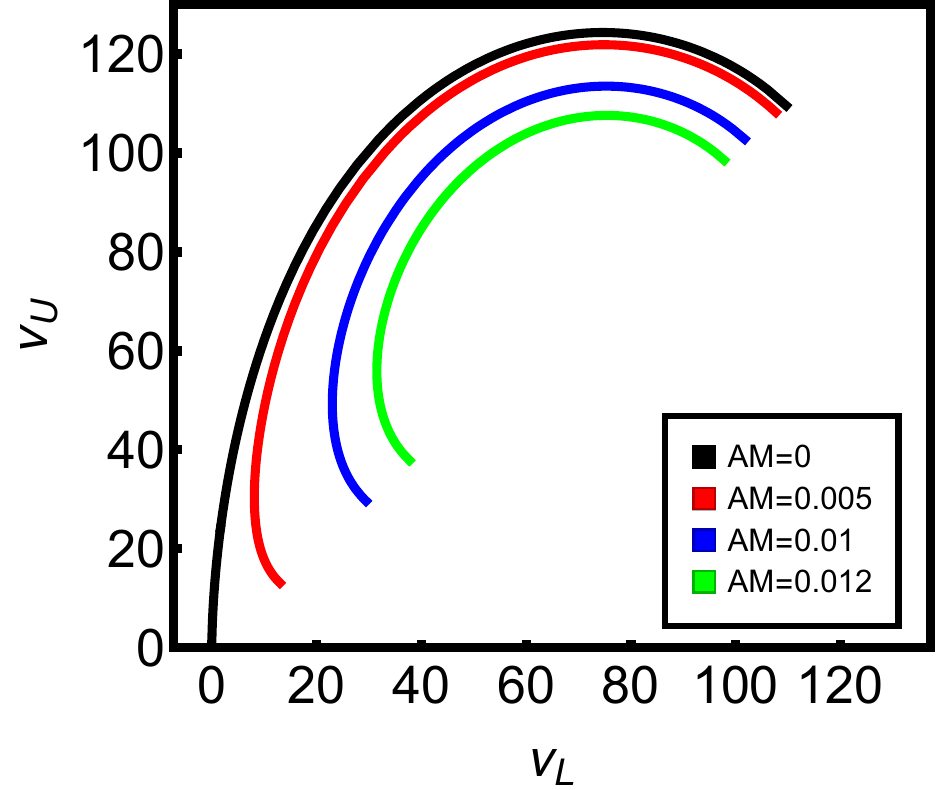}}
\subfigure[WD model]{\label{WD model}
\includegraphics[width=0.4\textwidth]{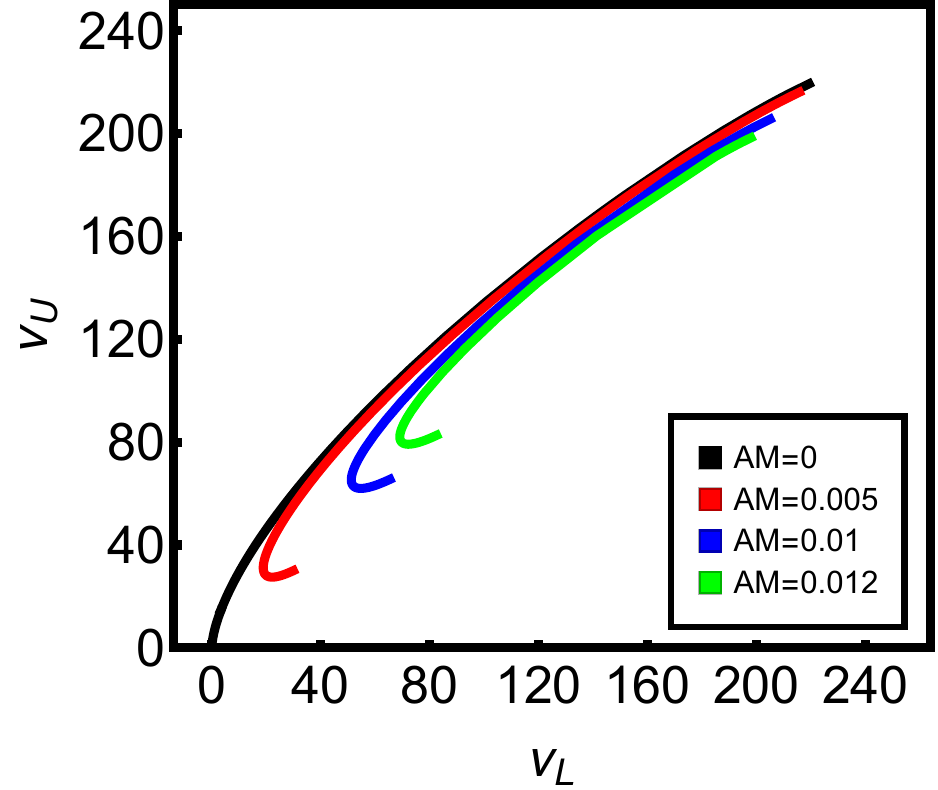}}
\caption{Relations between the upper ($\nu_U$) and lower ($\nu_L$) frequencies of twin-peak QPOs in the RP, ER3, ER4, and WD models around an accelerating Schwarzschild black hole, shown for various values of the acceleration parameter $AM$, assuming a black hole mass of $M=20M_\odot$.}\label{oscillation frequency}
\end{center}
\end{figure*}

Figure \ref{oscillation frequency} shows the relations between the upper and lower frequencies of twin-peak QPOs for different models. With the numerical calculations, for simplicity we set the mass of black hole with $M=20M_\odot$. From Fig. \ref{oscillation frequency}, we can see that increasing the acceleration parameter $A$ leads to a narrowing of both upper and lower frequency ranges across all models. This suggests that the acceleration factor $A$ has a suppressive effect on the QPO frequencies. Furthermore, since the domain of the radial epicyclic frequency $\Omega_r^2$ spans from the innermost to the outermost stable circular orbits—defined by the condition $\partial_r^2 V_{\text{eff}} = 0$, the upper and lower QPO frequencies coincide at these boundaries for all models, i.e., $\nu_U = \nu_L$. Notably, the WD model predicts systematically higher frequencies compared to the RP and ER models, indicating that warped thin disks may enhance QPO frequencies and thus serve as a viable explanation for higher-frequency QPO observations.

Furthermore, Abramowicz et al. proposed that twin-peak HF QPOs arise from resonances in nearly Keplerian accretion disk oscillations, implying that these QPOs should exhibit rational frequency ratios. Observational evidence supports this resonance hypothesis, particularly the $3:2$ ratio ($2\nu_U=3\nu_L$) commonly found in low-mass X-ray binaries (LMXBs) hosting black holes or microquasars \cite{Abramowicz:2002xc,Abramowicz:2004rr,Aliev:1980hz}.

\begin{figure}[htbp!]
\includegraphics[width=0.4\textwidth]{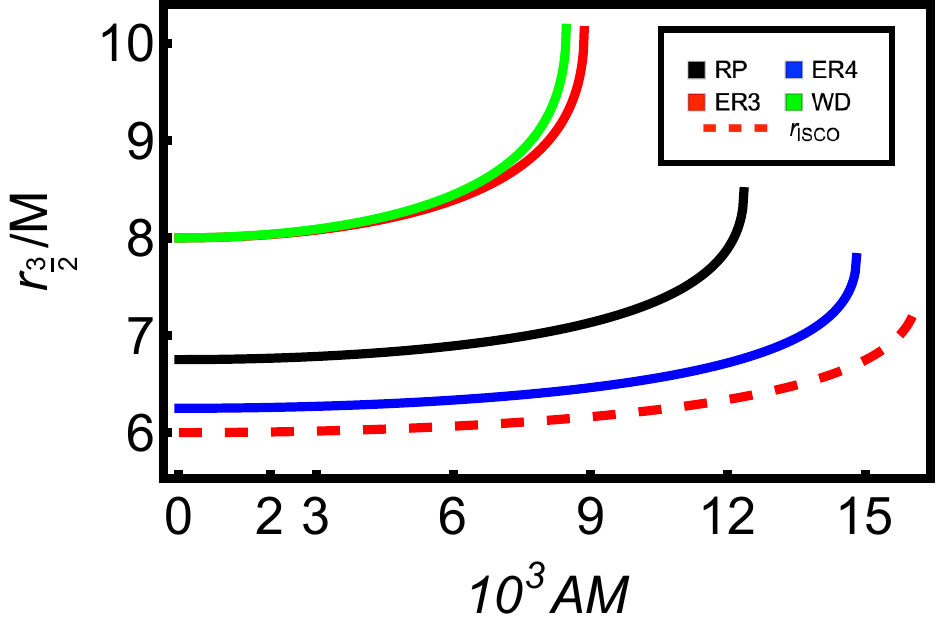}
\caption{Orbital radii corresponding to twin-peak QPOs with a $3:2$ frequency ratio and to the ISCO, plotted as functions of the acceleration parameter $AM$  for the RP, ER3, ER4, and WD models.}\label{radii of 32}
\end{figure}

By imposing the resonance condition $2\nu_U=3\nu_L$, we determine that the resonant radii $r_{3:2}$ corresponding to twin-peak HF QPOs with $2\nu_U=3\nu_L$ for varying acceleration factor $A$, and shown in Fig. \ref{radii of 32}. From Fig. \ref{radii of 32}, it is evident that the resonant radii $r_{3:2}$ and their dependence on acceleration factor $A$ are very similar between the ER3 and WD models, whereas the RP and ER4 models exhibit more distinct behavior. Moreover, compared to the other models, the resonant radii predicted by the ER4 model lie closest to the innermost stable circular orbit (ISCO).

\begin{figure}[htbp!]
\subfigure[$\nu_U$ with $r_{3:2}$]{\label{upper frequency}
\includegraphics[width=0.4\textwidth]{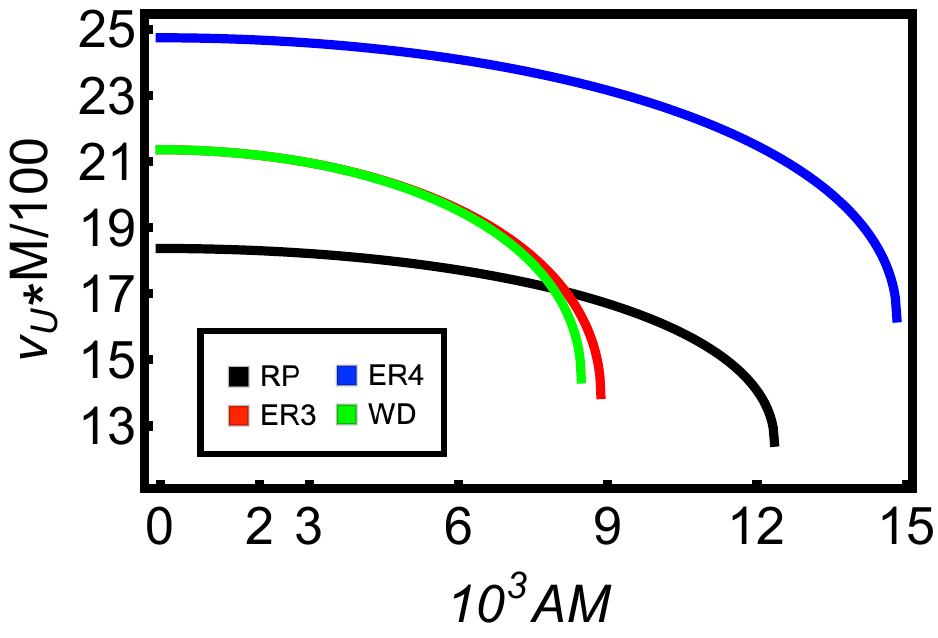}}
\subfigure[$\nu_{nod}$ with $r_{3:2}$]{\label{nodal precession frequency}
\includegraphics[width=0.4\textwidth]{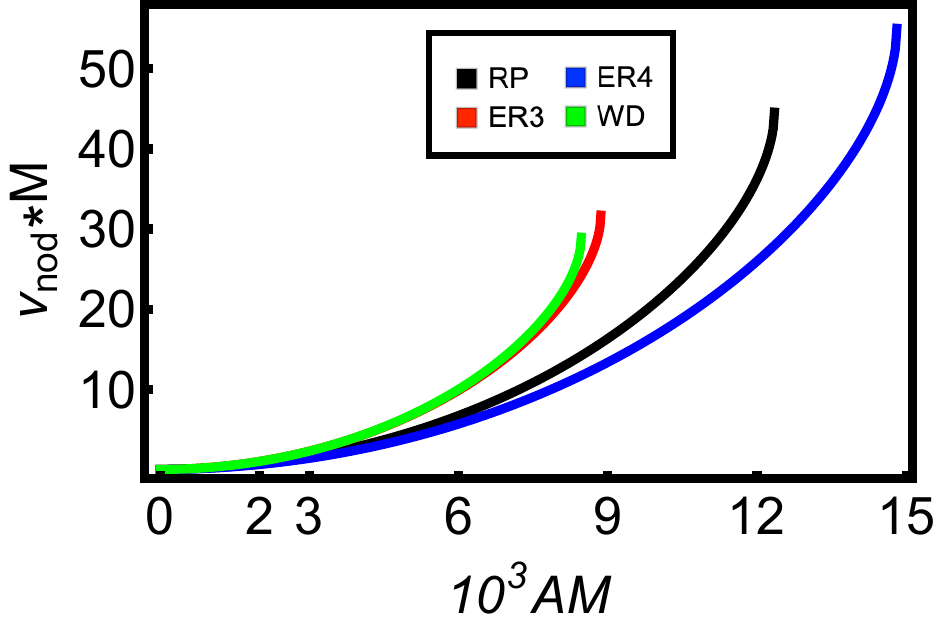}}
\caption{Upper frequency $\nu_U$ of twin-peak QPOs and the nodal precession frequency $\nu_{\text{nod}}=\nu_\phi-\nu_\theta$ at resonant radii where $\nu_U/\nu_L=3/2$, shown for the RP, ER3, ER4, and WD models.}\label{frequency of 32}
\end{figure}

Figure \ref{frequency of 32} depicts the upper QPO frequency $\nu_U$ and the nodal precession frequency $\nu_{nod}=\nu_{\phi}-\nu_{\theta}$ with the resonant radii $r_{3:2}$ for different QPO models. As shown in Fig. \ref{upper frequency}, the ER4 model explains higher QPO frequencies relative to the other three models. Although the resonant radius $r_{3:2}$ in the RP model is not the furthest from the ISCO, the RP model predicts comparatively lower QPO frequencies within the acceleration rang $AM*10^3\subset(0,8)$. From Fig. \ref{nodal precession frequency}, we can see that although the acceleration factor $A$ has a suppressive effect on the frequencies of twin-peak QPOs, the increase of acceleration factor $A$ can exert a reinforcement effect on nodal precession frequency $\nu_{nod}$. Furthermore, the acceleration factor $A$ demonstrates a more pronounced enhancement effect on nodal precession frequency $\nu_{nod}$ of the ER3 and WD models compared to the other two models. However, due to the greater span of the acceleration factor $A$ of ER4 model, this can result in a larger bandwidth of nodal precession frequency $\nu_{nod}$. 

\begin{figure}[htbp!]
\subfigure[RP and ER4 models for $r_{7:5}$, $r_{3:2}$, and $r_{8:5}$]{\label{radii of all a}
\includegraphics[width=0.4\textwidth]{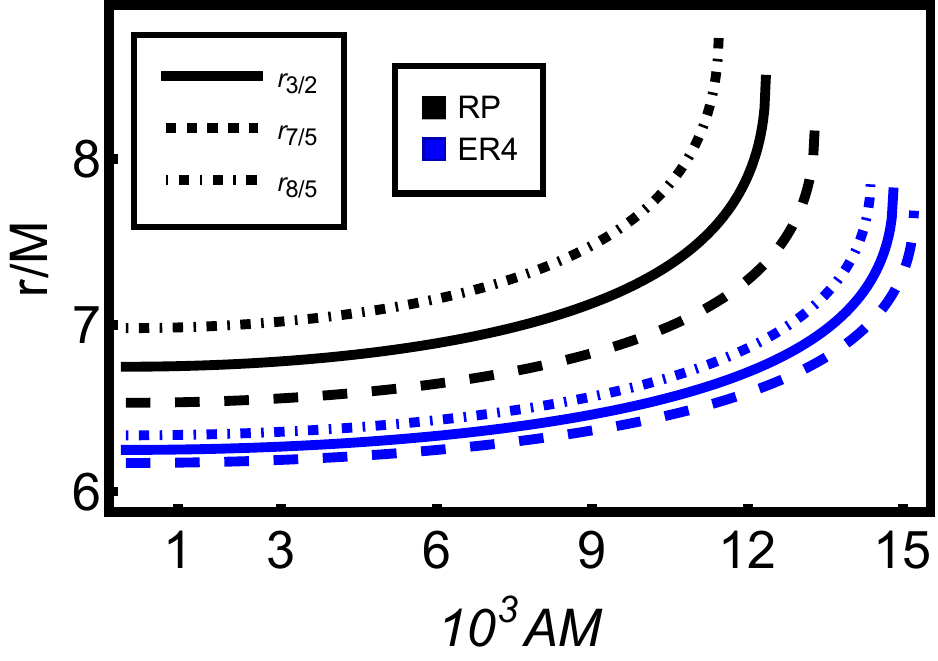}}
\subfigure[ER3 and WD models for $r_{7:5}$, $r_{3:2}$, and $r_{8:5}$]{\label{radii of all b}
\includegraphics[width=0.4\textwidth]{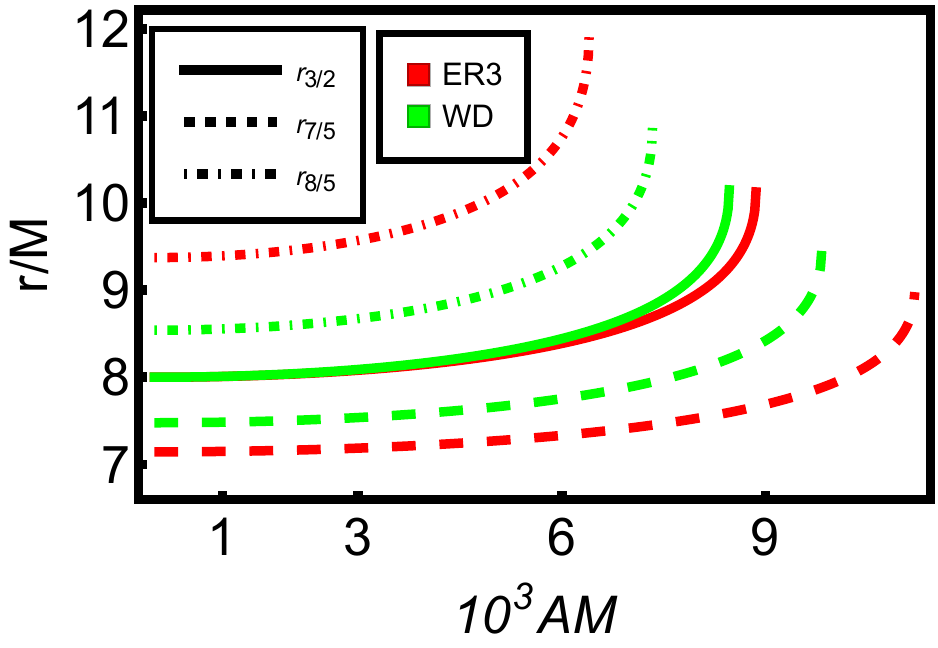}}
\caption{Orbital radii of twin-peak QPOs with frequency ratios $7:5$, $3:2$, and $8:5$ with different values of acceleration parameter $AM$, shown for the RP, ER3, ER4, and WD models.}\label{radii of all}
\end{figure}

Furthermore, we investigated other cases of different ratio in twin-peak HF QPOs with $5\nu_{U}=7\nu_L$ and $5\nu_{U}=8\nu_L$, with the corresponding resonant radii for different acceleration factors $A$ shown in Figure \ref{radii of all}. From this figure, it is clear that the resonant radius in the ER4 model remains nearly insensitive to the frequency ratio. In contrast, for the ER3 model, both the resonant radius and the allowed range of the acceleration factor $A$ are significantly influenced by the chosen frequency ratio.

Next, we constrain the acceleration factor $A$ by utilizing observational QPO data from two well-studied microquasars:\\
a): GRO J1655-40: Observations report upper and lower QPO frequencies of 441$\pm$2 Hz and 298$\pm$ 4 Hz, respectively, along with a nodal precession frequency  $\nu_{nod}$ with 17.3$\pm$0.1 Hz \cite{Motta:2013wga},\\
b): XTE J1859+226: This source exhibits upper and lower QPO frequencies of $227.5^{+2.1}_{-2.4}$ and $128.6^{+1.6}_{-1.8}$, respectively, with a nodal precession frequency $\nu_{nod}$ with $3.65\pm0.01$ \cite{Motta:2022rku}.

\begin{figure*}[!htb]
\begin{center}
\subfigure[GRO J1655-40]{\label{J1655 ER4 model}
\includegraphics[width=0.4\textwidth]{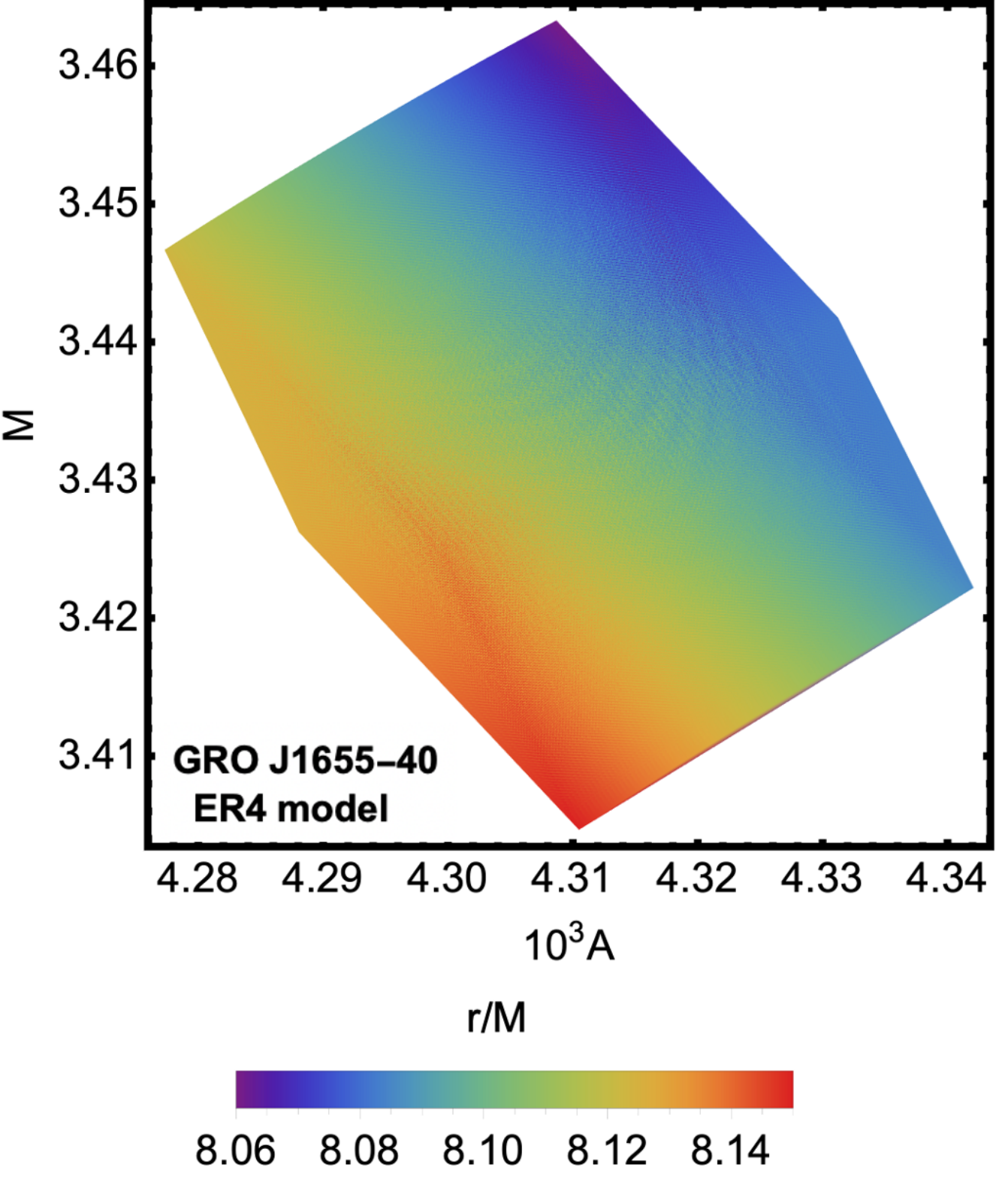}}
\subfigure[XTE J1859+226]{\label{J1859 RP}
\includegraphics[width=0.4\textwidth]{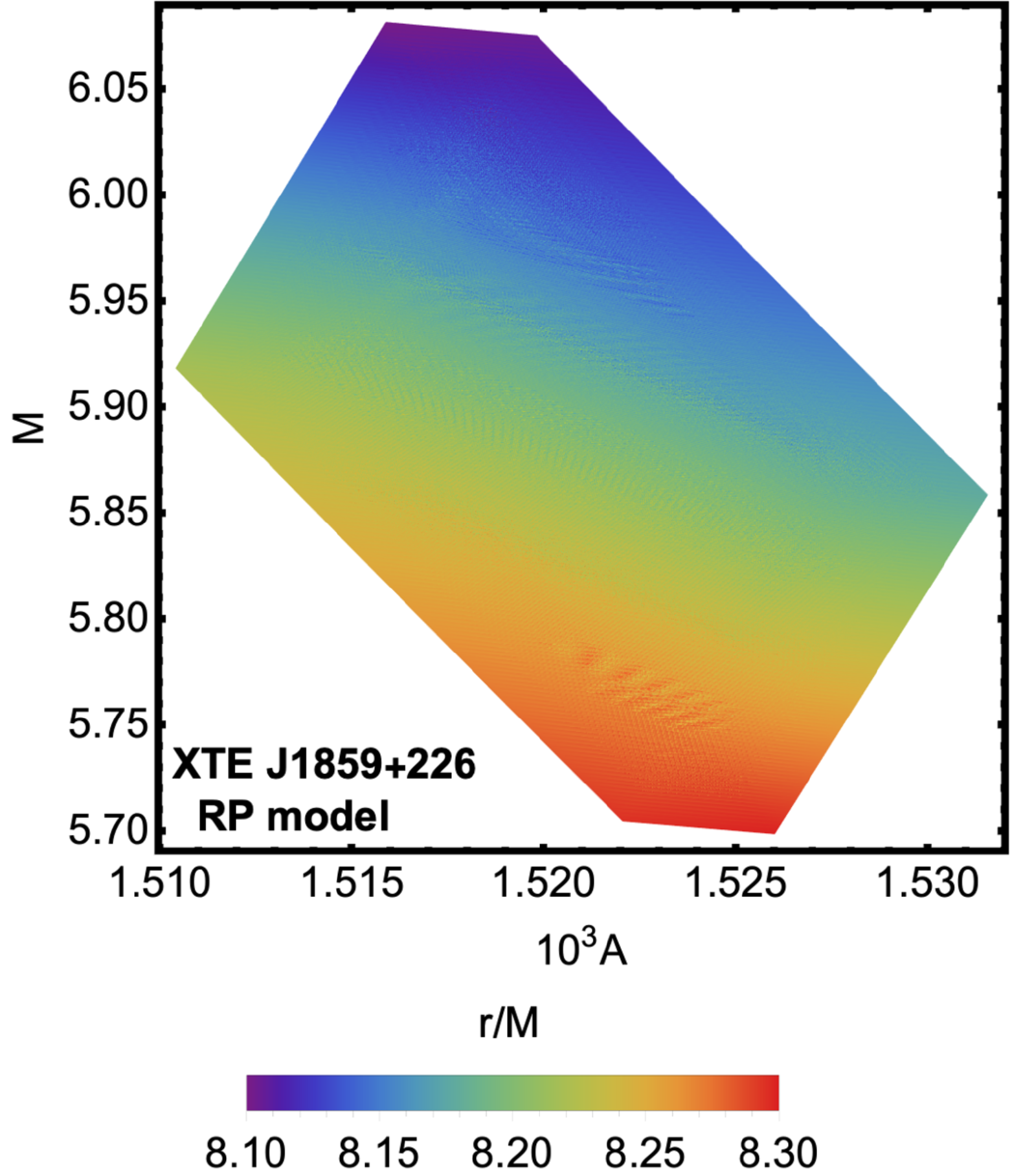}}
\subfigure[XTE J1859+226]{\label{J1859 ER4}
\includegraphics[width=0.4\textwidth]{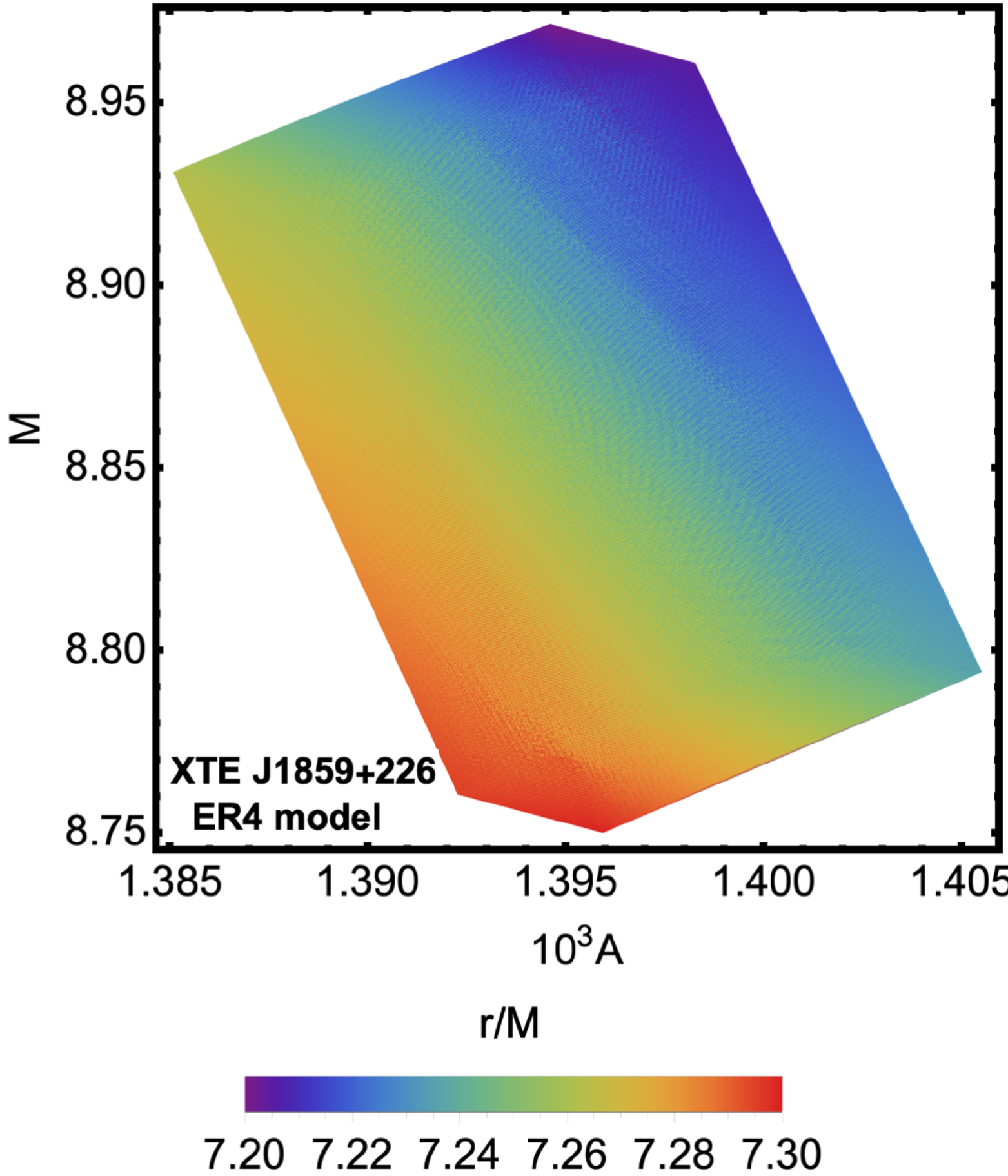}}
\subfigure[XTE J1859+226]{\label{J1859 WD}
\includegraphics[width=0.4\textwidth]{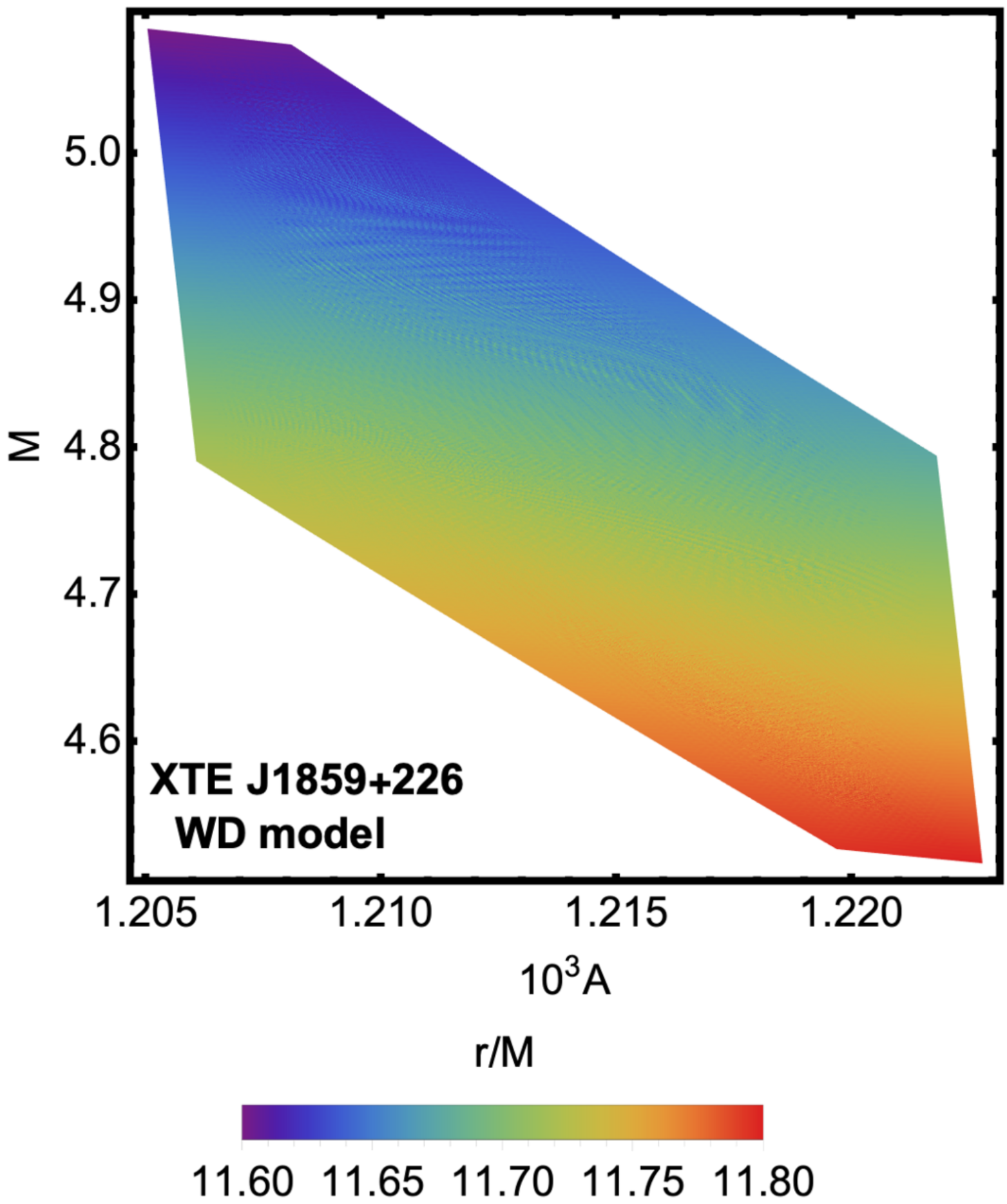}}
\caption{Constraints on the acceleration factor $A$, black hole mass $M$, and normalized QPO orbit radius $r/M$ derived from observational QPO data and TOV limits for the stellar-mass black holes GRO J1655–40 and XTE J1859+226.}\label{data constrain}
\end{center}
\end{figure*}

With a straightforward numerical calculation 
\begin{eqnarray}
\nu_{i}(A,M,r)=\nu_{i}^{obs},~~(i=U,L,nod),
\end{eqnarray} 
and combining the Tolman–Oppenheimer–Volkoff (TOV) limit, which can infer that the mass of the black hole should exceed $3M_\odot$, we can get the rang of corresponding parameters (A, M, r) for the accelerating black hole, which is depicted in Fig. \ref{data constrain}.

From Fig. \ref{J1655 ER4 model}, we can see that, due to the microquasars GRO J1655-40 with a relatively observed higher frequency of twin-peak QPOs, the TOV limit can impose a stringent constraint on potential models for explaining the QPO around the accelerating black hole, and delineate a narrow spectrum of parameters $(10^3A, M, r/M)$ with $(4.31, 3.43M_\odot, 8.08)$, within which only the ER4 model provides a consistent explanation. For the microquasars XTE J1859+226 case, shown in Fig.\ref{J1859 RP}-\ref{J1859 WD}, the TOV limit excludes the ER3 model as a viable interpretation of the observational data. Among the other three remaining models, the ER4 model predicts the largest black hole mass estimate, $(M=8.8M_\odot)$, with QPOs originating at radii closest to the innermost stable circular orbit. In contrast, the WD model yields the smallest mass estimate, $(M=4.8M_\odot)$, with QPO generation occurring furthest from the event horizon.  Furthermore, for the XTE J1859+226 case, the acceleration factor $A$ values estimated by the three interpretable models congregate around $10^3A\approx1.4$. However, in comparison to the GRO J1655-40 case with $10^3A\approx4.3$, it can be inferred that the microquasars GRO J1655-40 is situated within a superior acceleration factor.

\section{Conclusion and discussion}\label{conclusion}

{In this paper, we have investigated the dynamics of massive test particles in the spacetime of an accelerating black hole and analyzed the characteristics of quasi-periodic oscillations (QPOs) associated with such backgrounds.

Our analysis demonstrates that the acceleration factor $A$ suppresses the radial effective potential $V_{\text{eff}}$, thereby reducing both the total energy $\mathcal{E}$ and angular momentum $\mathcal{L}$ required for circular orbits. The condition for the existence of stable circular orbits, $\partial_r^2 V_{\text{eff}} \geq 0$, constrains the acceleration parameter to $A M \leq 0.0161$. The radius of the ISCO grows, and this growth rate accelerates with larger acceleration factor $A$. In contrast, the angular momentum $\mathcal{L}_{\text{ISCO}}$ and energy $\mathcal{E}_{\text{ISCO}}$ associated with the ISCO both decrease. Furthermore, the radiative efficiency $\epsilon$ increases monotonically with $A$, reaching a maximum value of approximately $6.9\%$. 

We further examined the fundamental frequencies of test particles orbiting the accelerating black hole. The acceleration factor $A$ enhances the decay of the Keplerian frequency $\Omega_{\phi}$, however since $\Omega_{\phi}$ remains monotonic, the Aschenbach effect is absent. The vertical frequency $\Omega_{\theta}$ exhibits behavior similar to $\Omega_{\phi}$, but the difference between $\Omega_{\phi}$ and $\Omega_{\theta}$ increases with $A$, distinguishing accelerating black holes from spherically symmetric ones. The radial frequency $\Omega_r$ is confined to the region of stable circular orbits and is suppressed in both magnitude and radial extent by the acceleration factor.

Finally, we analyzed twin-peak high-frequency QPOs (HF QPOs) with the framework of four representative models: the relativistic precession (RP), epicyclic resonance (ER3 and ER4), and warped disc (WD) models. Among these, the WD model, which assumes a warped thin accretion disk, predicts systematically the highest QPO frequencies. While the resonant radii in the ER4 model are largely insensitive to the fixed frequency ratio of twin-peak HF QPOs, the ER3 model shows significant sensitivity to this ratio. Although the acceleration factor $A$ suppresses the frequencies of twin-peak QPOs overall, it enhances the nodal precession frequency $\nu_{\text{nod}}$. By fitting observational data from GRO J1655-40 and XTE J1859+226, combined with the Tolman–Oppenheimer–Volkoff (TOV) limit, we find that only the ER4 model provides a consistent explanation for GRO J1655-40 with parameters $(10^3 A, M, r/M) \approx (4.31, 3.43 M_\odot, 8.08)$. For XTE J1859+226, the TOV limit excludes the ER3 model, while the remaining models estimate the acceleration factor around $10^3 A \approx 1.4$. These results suggests that GRO J1655-40 experiences a more pronounced acceleration effect than XTE J1859+226.
}

\begin{acknowledgments}
{This work is supported in part by the Anhui University of Science and Technology under Grant No. (2024yjrc164 and YJ20240001)
}.
\end{acknowledgments}


\end{document}